\documentclass[preprint,prl,aps]{revtex4} %
\usepackage{graphicx}
\usepackage{dcolumn}
\usepackage{bm}
\usepackage{epsf}
\begin{document}
\title{Observation and theoretical description of the pure 
Fano-effect in the valence-band photo-emission of ferromagnets}
\author{J.\ Min\'ar and H.\ Ebert} 
\address{Department Chemie,
  Physikalische Chemie, Universit\"at M\"unchen, Butenandtstr. 5-13,
  D-81377 M\"unchen, Germany}
\author{C.\ De Nada\"i, N.\ B.\ Brookes and F.\ Venturini}
\address{European Synchrotron Radiation Facility, Bo\^{i}te Postale
  220, 38043 Grenoble Cedex, France}
\author{G.\ Ghiringhelli}
\address{INFM- Dip. di Fisica, Politecnico di Milano, p. Leonardo da
  Vinci 32, 20133 Milano, Italy}
\author{L. Chioncel$^*$ and  M. I. Katsnelson}
\address{University of Nijmegen,
NL-6525 ED Nijmegen, The Netherlands\\
$^*$Institut f\"ur Theoretische Physik - Computational Physics,
Technische Universit\"at Graz, A-8010 Graz, Austria}
\author{A. I. Lichtenstein}
\address{Institut f\"ur Theoretische Physik, 
Universit\"at Hamburg, 20355 Hamburg, Germany}

\date{\today} 
\begin{abstract}
The pure Fano-effect in angle-integrated valence-band
 photo-emission of ferromagnets has been observed for 
the first time. A contribution of the intrinsic spin 
polarization to the spin polarization of the 
photo-electrons has been avoided by an appropriate choice of the
experimental parameters. The
 theoretical description of the resulting spectra
 reveals a complete  analogy to the Fano-effect observed
 before for paramagnetic transition metals. While the 
theoretical photo-current and spin difference spectra
 are found in good quantitative agreement with 
experiment in the case of Fe and Co only a qualitative
 agreement could be achieved 
 in the case of Ni by calculations on the
 basis of plain local spin density approximation (LSDA).
Agreement with experimental data could be improved 
in this case 
in a very substantial way
by a  treatment  of      
correlation effects 
on the basis of  dynamical mean field theory (DMFT). 
\end{abstract}

\draft \preprint{IP/1}


\narrowtext

\maketitle

%
Spin-resolved photo-emission spectroscopy
is a powerful tool to study the magnetic
aspects of the electronic structure of ferromagnetic materials. This
could be demonstrated already about 25 years ago by the pioneering
work of Eib and Alvarado\cite{EA76} who investigated for Ni(110) the
spin polarization of the photo-electrons in a photo-threshold
experiment. More detailed information is obtained from
spin-, energy- and angle-resolved photo-emission experiments that became
feasible by the use of samples with a remanent magnetization oriented
perpendicular to the electron-optical axis.\cite{KGK+80}
This technique was in particular used to study the dependence of the
electronic and magnetic properties of the 3d-ferromagnets on
temperature.\cite{KSCG84} Even more refined experiments became
possible by the use of circularly polarized radiation, that allows to
study magnetic circular dichroism. As it could be demonstrated
theoretically\cite{SHHF94,Bra96} 
and experimentally,\cite{KS01}
spin-resolved photo-emission experiments using circularly polarized
radiation allow in particular to reveal the hybridisation of states
with different spin character due to spin-orbit coupling.

Magnetic circular dichroism in magnetically ordered systems is closely
related to the Fano-effect\cite{Fan69a} that also occurs as a
consequence of the spin-orbit coupling. The term Fano-effect denotes
the observation that one can have a spin-polarized photo-current from
a non-magnetic sample if circularly polarized radiation is used for
excitation. While for a non-magnetic sample the spin-polarization of
the photo-current is reversed if the helicity of the radiation is
reversed, this symmetry is in general broken for a magnetically
ordered system leading to magnetic circular dichroism. This implies in
particular that if a spin-resolved photo-emission experiment is done
with circularly polarized radiation coming in along the direction of the
magnetization of a ferromagnetic material and spin analysis of the
photo-current is done with respect to this direction, the
spin-polarization of the photo-current due to spin-orbit coupling is
superimposed to that due to magnetic ordering. In the following it is
demonstrated by experiments on Fe, Co and Ni
 that the pure Fano-effect can also be observed in
angle-integrated valence band X-ray photo-emission spectroscopy
(VB-XPS) for ferromagnets, if the
circularly polarized 
 radiation impinges perpendicular to the magnetization and
if subsequent spin analysis is done with respect to the direction of
the photon beam.

Accompanying calculations based on local spin-density approximation
(LSDA) and using a fully relativistic implementation of the one-step
model of photo-emission allow for a detailed discussion of the
experimental spectra. 
As found before by comparable VB-XPS investigations,
calculations based on plain LSDA lead for Ni
 to a band width that
seems to be too large compared with experiment and are
not able to
reproduce the satellite structure observed at 
6~eV binding energy.\cite{HW72,HW75}
These short comings have been ascribed to an
inadequate treatment of correlation effects within LSDA. To remove
these problems several approaches have been used in the
 past.\cite{Lie79,Ary92,IUHF94,MBO+99} 
Here we show that the use of LSDA in combination with 
dynamical mean field theory
(DMFT) leads for Ni to a substantial
improvement of the agreement of 
theoretical and experimental VB-XPS spectra.

 The experiments were performed on the helical undulator beamline ID08
 at the ESRF (European Synchrotron Radiation Facility). The APPLE II
 insertion device provides approximately 100$\%$ polarised soft X-rays 
(vertical and horizontal linearly polarised and left and right circularly
polarised). All experiments have been performed on thin films
($2-4$~nm) prepared in-situ by e-beam growth on a clean
Cu(001) substrate. The base pressure in the chamber was in the $10^{-11}$~mbar 
range.
In all cases the shape anisotropy gave rise to an orientation of the
magnetisation in-plane with a multidomain structure. The magnetisation
of the sample was measured in remanence using X-ray magnetic circular
dichroism. In the measuring geometry of normal incidence a vanishing
dichroism signal was observed. 
The spin polarised photoemission measurements were performed at room
temperature with the X-rays normal to the sample and the electrons
were analysed with a hemispherical electron energy analyser with a $\pm
20^\circ$ angular acceptance at $60^\circ$ 
to the incident light. Together with the polycrystalline nature of the
samples this means that the measurement is angle integrated.
The combined energy resolution was $\approx 0.5
- 0.7$~eV. The spin of the electrons was measured using a mini-Mott
spin detector after the energy analysis \cite{GLB99}. In order to eliminate
instrumental asymmetries the spectra were measured with both
helicities of circular polarised light. This allows one to extract a
spectra that is equivalent to the calculated spectra with a single
helicity. The
experimental spin polarisation is also corrected for the measuring
geometry to give the spin polarisation along the light propagation
direction. 


To deal with the geometry of the photo-emission experiment 
described above we adopt the spin-density matrix formalism 
as described for example by Ackermann and Feder\cite{AF85a}
for the angle-integrated case.
This approach allows to express the photo-current and its spin
polarization in terms of the spin density matrix
\begin{eqnarray}
\rho (E')&=&\sum_{m_s m_s\;'}
|m_s\rangle
 \tilde{I}_{m_s m_{s\;'}} 
\langle m_{s\;'}|
\; ,
\end{eqnarray}
with the angle-integrated spin dependent intensity function
\begin{eqnarray}
\label{Eq:Iss}
\tilde{I}_{m_s m_{s'}}^{\lambda}
     &=& - \frac{1}{\pi} \Im \int d\hat{k}  \\
&& \quad <\Psi_{m_s
  \vec k}^{final}(E')|
X_{\vec q \lambda} G(E) X_{\vec q \lambda}^\dagger
|\Psi_{m_s' \vec k}^{final}(E')>  \nonumber \,.
\end{eqnarray}
Here the initial valence band states at energy $E$ are represented by
the single particle Green's functions $G(E)$ and absorption of radiation
with wave vector $\vec q$, frequency $\omega$ and polarization
$\lambda$ represented by the electron-photon interaction operator
$X_{\vec q \lambda}$\cite{Ros61} is considered. The final states
at energy $E'=E+\hbar \omega$ are given by a time-reversed
spin-polarized LEED state.\cite{AF85a}
With the spin density matrix $\rho$ available the spin-averaged
photo-current intensity $\bar I$ and photo-electron spin polarization
$\vec P$ are given as $\bar I = {\rm Trace} \, \rho$ and
$\vec P = {\rm Trace}({\vec \sigma} \rho)/\bar I$, respectively,
with $\vec \sigma $ the vector of spin matrices.
Finally, if a spin analysis of the photo-current is performed with
respect to a direction $\hat n$, the corresponding spin-projected
photo-current $I_{\sigma}$ is given by:
\begin{eqnarray}\label{Eq:I-sig}
I_{\sigma}^{\lambda} &=& (1+\sigma \vec P \cdot \hat n) \bar I/2 
\end{eqnarray}
with $\sigma = \pm 1$ corresponding to spin-up and
spin-down.
In the following, the spin density matrix is 
defined with respect to a right-handed coordinate
 system with its z-axis chosen along the magnetization 
of the sample that in turn is oriented parallel to the 
surface plane. The x-axis coincides with the surface 
normal that specifies the direction
$\hat n$  for the 
spin analysis. 
To calculate the angle integrated spin-dependent intensity function
 $\tilde{I}_{m_s m_{s'}}$ given by Eq.\ (\ref{Eq:Iss}) 
for  a  photo-emission experiment using
 circularly polarized radiation an appropriate
 extension of the fully relativistic approach 
worked out by Ebert and Schwitalla\cite{ES97} has been made.
This implies first of all that electronic properties are obtained within
the framework of LSDA. To calculate the
initial state Green's function and also the involved final states multiple
scattering theory is used.\cite{Ebe00} Because of the relatively high
photon energy used in the experiments described below, the
single-scatterer approximation has been used for the later ones.
Finally, to compare theoretical spectra based on Eq.\ (\ref{Eq:I-sig})
various broadening mechanisms are incorporated in a phenomenological
way.
Intrinsic life-time effects are described by a
 Lorentzian-broadening with an energy dependent
 width $\Gamma(E) = a + b(E-E_F)^2$ with $E_F$ the Fermi
 energy. Instrumental broadening in turn is accounted by
 Gaussian broadening with a broadening parameter $\sigma$.
 (For the spectra to be shown below the following
 parameters have been used:
 $a=0.01$~eV, $b=0.01$~eV$^{-1}$, $\sigma=0.4$~eV ).

When dealing with the Fano-effect in 
paramagnetic noble metals\cite{MEG+01,NME+04} 
and in the ferromagnets Fe and Co (see
below), the approach sketched above lead to theoretical spectra in
rather good agreement with experiment. In the case of Ni, however,
pronounced deviations occur that have to be ascribed to correlation
effects that are not accounted for in an adequate way by LSDA. To
remove these problems, we applied a recently proposed self-consistent 
KKR+DMFT scheme.\cite{MCP+05} 
Within the relativistic extension of
this approach, correlation effects are represented
by a complex and local self-energy $\Sigma (E)$ that enters the
Dirac-Hamiltonian used to calculate the Green's function 
$G(E)$ in
Eq.~(\ref{Eq:Iss}). To calculate in turn $\Sigma(E)$, the most
general rotationally invariant form  of the
generalized Hubbard (on-site) Hamiltonian\cite{AAL97}
has been taken for the
interaction Hamiltonian.
Within  DMFT, the
 many-body problem for a crystal is split into a one-particle
impurity problem for the crystal and a many-body problem for one site in
an effective medium
 (for the effective impurity method see for example
Refs.\ \onlinecite{GKKR96,KL99}).
 The
 correlation effects are treated in the framework of dynamical mean
 field theory (DMFT) \cite{GKKR96}, with a spin-polarized
 T-matrix Fluctuation Exchange (SPTF) type of DMFT solver
 \cite{KL02}. The SPTF approximation is a multiband
 spin-polarized generalization of the fluctuation exchange
 approximation (FLEX) \cite{BS89a,KL99}. Since some part of
 the correlation effects are included already in
the local spin-density approximation (LSDA) ``double counted''
terms should be taken into account. To this aim, we start with the
LSDA electronic structure and replace $\Sigma_{\sigma}(E)$ by
$\Sigma_{\sigma}(E)-\Sigma_{\sigma}(0)$ in all equations of the
LSDA+DMFT method \cite{LKK01}, the energy $E$ being relative to the
Fermi energy and $E_F=0$. It means that we only add {\it
dynamical} correlation effects to the LSDA method.

%

\begin{figure}[htbp]
\leavevmode
\includegraphics[height=7cm]{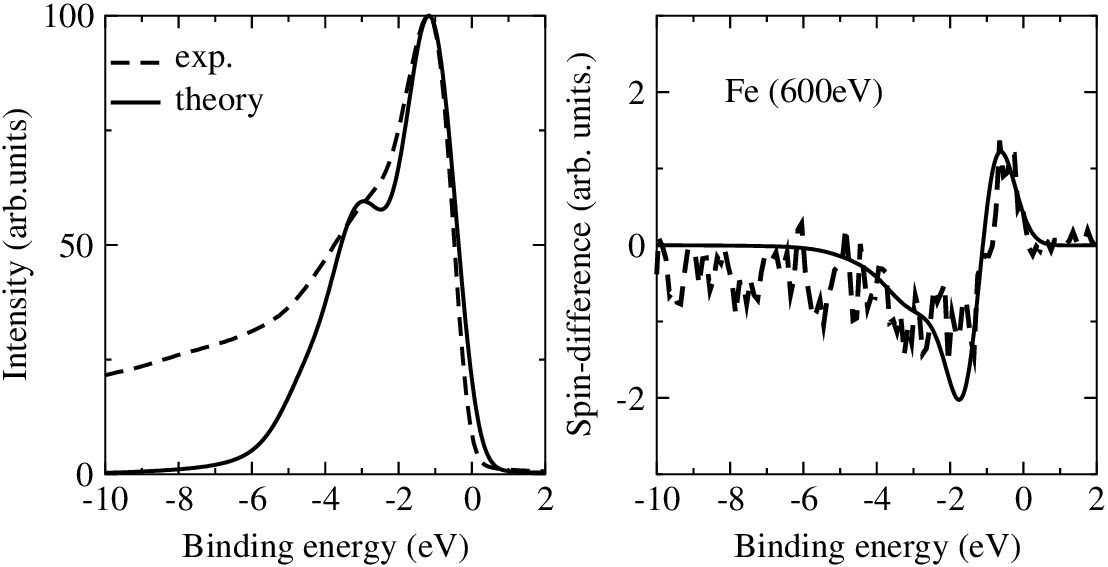}
\includegraphics[height=7cm]{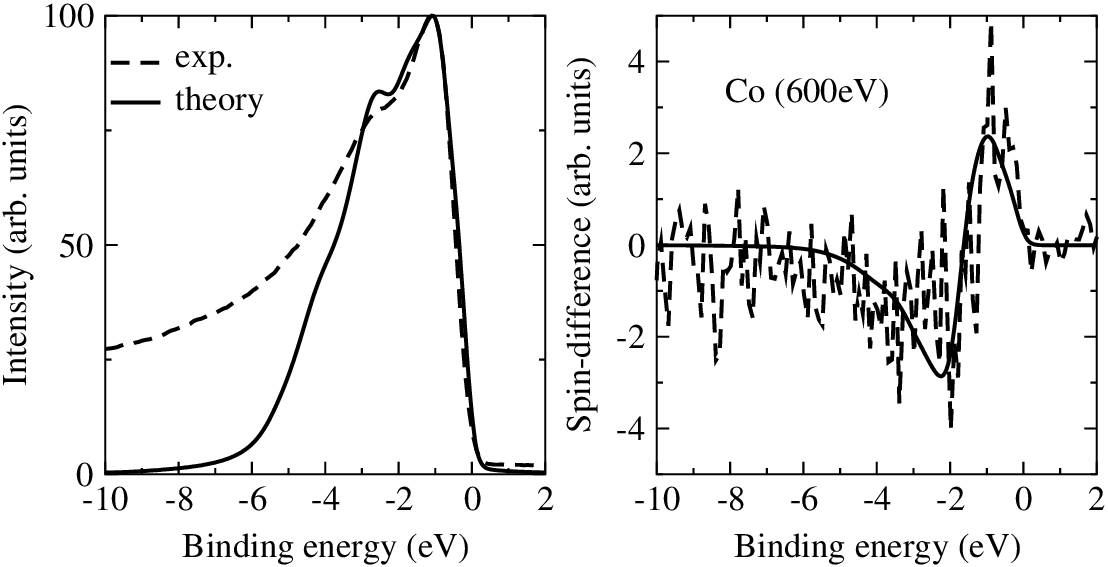}
\includegraphics[height=7cm]{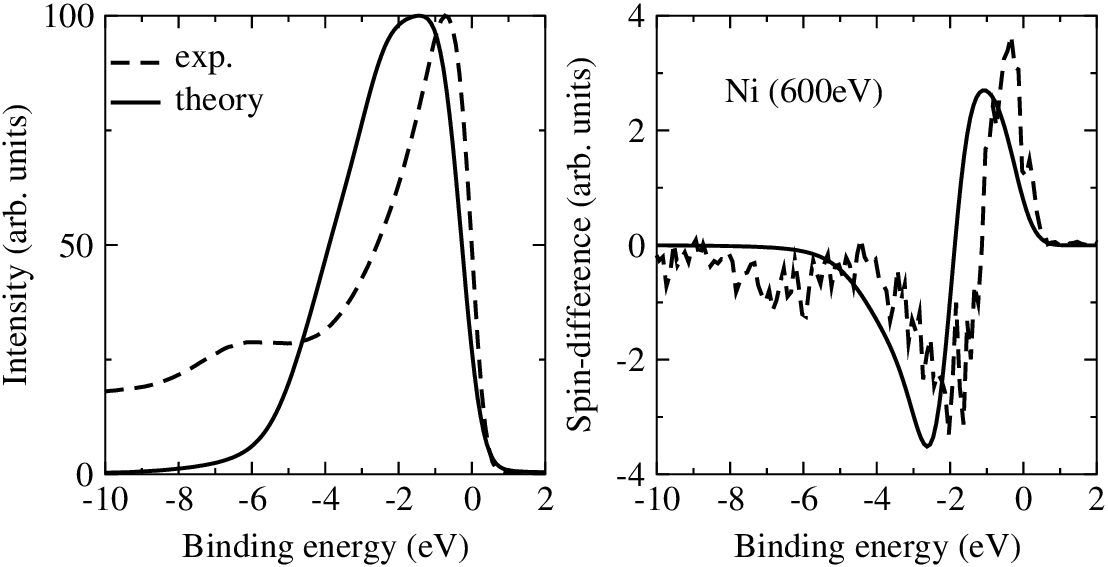}
\vspace{0.5cm}

\caption{Left: spin and angle-integrated VB-XPS-spectrum of 
  ferromagnetic Fe, Co and Ni for a photon energy of 600~eV.  Right:
  spin-difference 
  $\Delta I^+=I^+_{\uparrow}-I^+_\downarrow$ of the photo-current for
  excitation with left circularly polarized radiation. Theory: full
  line; experiment: dashed line. The same scale has been used for the
  intensity and corresponding spin-difference plots.}
\label{FIG:IXPS}
\end{figure}

\begin{figure}[htbp]
\includegraphics[height=8cm]{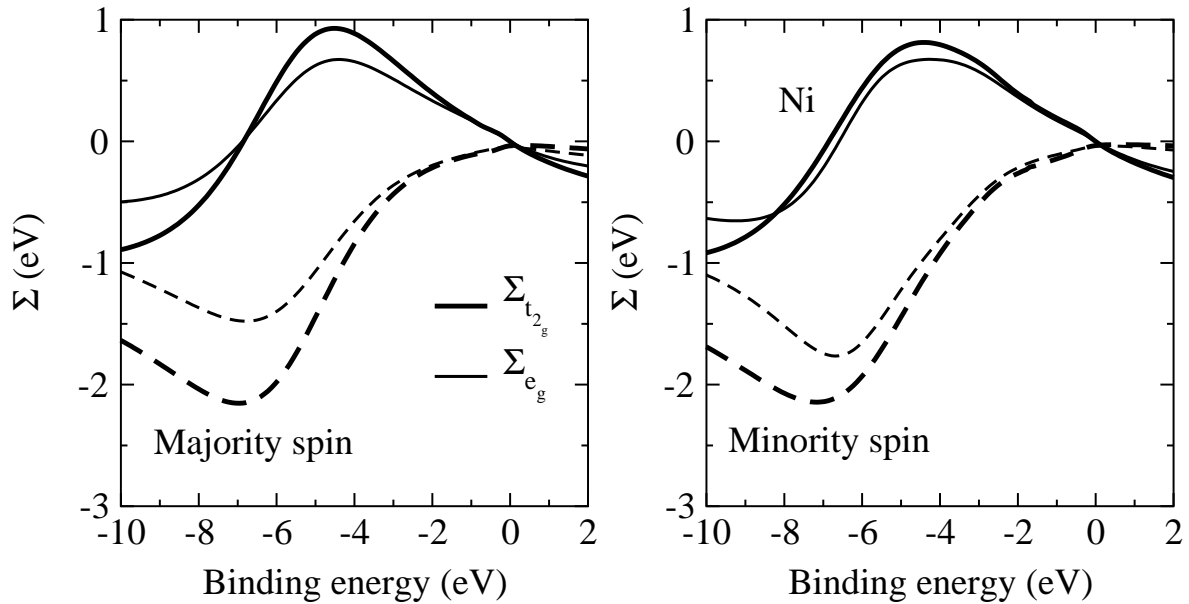}
\vspace{0.5cm}

\caption{The spin resolved self-energies of ferromagnetic Ni
  calculated within the 
  DMFT for U=3~eV, J=0.9~eV and T=500K. The real parts (full lines) and
  imaginary parts (dashed lines) are shown separately for the 
  t$_{2_{g}}$(thick lines) and e$_{g}$(thin lines) representations
  of the d-states.}
\label{FIG:SIGMA}
\end{figure}

\begin{figure}[htbp]
\leavevmode
\includegraphics[height=8cm]{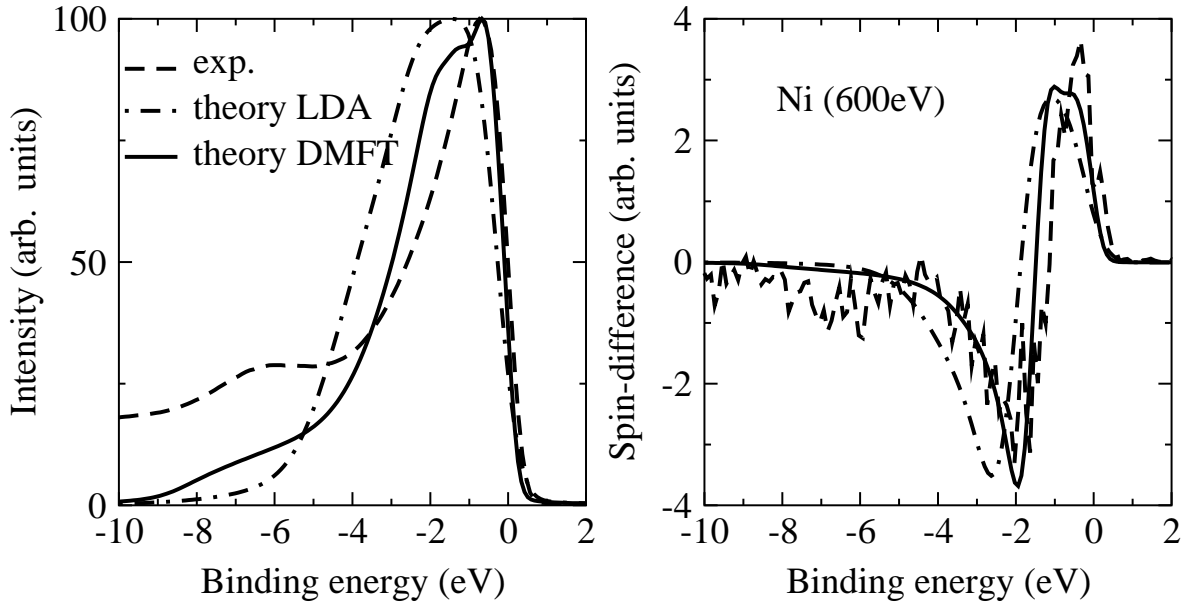}
\vspace{0.5cm}

\caption{Left: The experimental (dashed lines), theoretical LSDA (dot
  dashed lines) and LSDA+DMFT (full line)
  data for spin and angle integrated VB-XPS spectra of 
 Ni for a photon energy of 600~eV.  Right: spin-difference
   $\Delta I^+=I^+_{\uparrow}-I^+_\downarrow$  of the photo-current for
  excitation with left circularly polarized radiation.}
\label{FIG:NISELF}
\end{figure}

The experimental VB-XPS spectra of Fe, Co- and Ni 
recorded for a photon-energy of 600~eV and normalized 
to a peak height of 100 are shown on the left side in
 Fig.~\ref{FIG:IXPS}.
%
%
%
 No subtraction of the secondary background has 
been made. Comparison of the spectra with those of 
previous experimental work leads to a fairly good 
agreement. The right hand side of Fig.~\ref{FIG:IXPS} gives the same
scale the 
observed spin difference of the photo-current 
$\Delta I^+$, i.e.\ the difference of the currents of 
photo-electrons with spin-up and spin-down,
for an excitation with left circularly polarized radiation. Because 
the polarisation analysis of the photo-current is done with 
respect to an axis that is perpendicular to the 
spontaneous magnetization $\vec{m}$, $\Delta I^+$ cannot 
be caused by the exchange splitting of the ground state. 
In fact, one finds that the shape and amplitude of the 
$\Delta I^+$ curves are very similar to that found for 
paramagnetic Cu.\cite{MEG+01} This finding suggests 
that the observed spin difference reflects in both 
cases the Fano-effect that is due to the presence of 
spin-orbit coupling and the use of circularly polarized 
radiation for excitation. This can indeed be confirmed 
by an extension of the analytical model developed 
recently when dealing with the Fano-effect in the VB-XPS
 of paramagnetic solids.\cite{MEG+01} To support this 
interpretation of the $\Delta I^+$ spectra ab initio calculations 
have been performed along the lines sketched above. Taking into
 account the influence of the secondary electrons our theoretical
 results for the VB-XPS spectra are in fairly good agreement with
 experiment in the case of Fe and Co. For Ni, on the other hand, 
the LSDA-based calculations lead to a band-width that is much too 
large. Furthermore they are not able to reproduce the satellite 
at about 6~eV binding energy. To deal with this well-known problem 
connected with the valence-band photo-emission of Ni, additional 
calculations have been made on the basis of the LSDA+DMFT scheme 
(for the corresponding parameters see caption of 
Fig.~\ref{FIG:SIGMA}). 
The 
resulting self-consistent complex and energy dependent self-energy $\Sigma(E)$ 
is shown in Fig.~\ref{FIG:SIGMA} in a spin- and symmetry-resolved way. 
%
%
%
%
The appreciable real part of $\Sigma(E)$ gives rise to a 
corresponding shrinking of the d-band width of Ni. This
 leads to a much better agreement of the theoretical
 VB-XPS spectrum with experiment, as can be seen in 
Fig.~\ref{FIG:NISELF}. 
%
%
In addition, use of the LSDA+DMFT scheme leads
 to a pronounced increase of the intensity in the 
regime of the 6~eV satellite.

The theoretical spin difference $\Delta I^+$ 
shown on the right hand side of Fig.~\ref{FIG:IXPS}
is found in rather good agreement with experiment
in particular for Fe and Co.
As mentioned, the shape and amplitude of the curves are
 very similar to that found for paramagnetic Cu. 
In fact, performing model calculations for Fe, Co and Ni
 with the exchange splitting suppressed, i.e.\ for a 
hypothetical paramagnetic state, one obtains $\Delta I^+$ 
curves that differ not very much from those for the 
ferromagnetic state (in particular for Fe and Co the
 curves differs somewhat with respect to shape and amplitude).
 The sequence for the maximum (minimum) of $\Delta I^+$ of Fe, Co and 
Ni is found to be 1.3, 2.2 and 2.6 (-1.9, -2.7 and -3.5).
 Although there are several electronic and structural 
properties that determine these data, they nevertheless
 correlate reasonably well with the strength of the 
spin-orbit coupling parameters of the d-states 
(66, 85 and 107~meV, respectively)\cite{PEND01} 
to identify once more the spin-orbit coupling 
as the source for the observed spin-polarization.
Performing calculations for a reversed helicity 
$\lambda$ of the radiation leads to a reversed sign for the
spin polarisation   $\Delta I^{\lambda}$.
This is an additional proof that the pure Fano-effect
is indeed observed by the experimental set-up described above.
As for the standard VB-XPS spectra inclusion of the 
self-energy $\Sigma(E)$ leads to a substantial
 improvement for the agreement of the 
theoretical $\Delta I^+$ spectrum with experiment.
 As one can see in Fig.~\ref{FIG:NISELF} the shrinking of the 
band width is also reflected by the $\Delta I^+$ curves, 
while their amplitude and shape is only slightly changed.

%
In conclusion, it has been demonstrated that the pure Fano-effect 
in angle-integrated VB-XPS can also be observed in
 ferromagnets if an appropriate geometry is chosen.
 The interpretation of the resulting spin polarization
 spectra can be done in complete analogy as for the 
paramagnetic case. Corresponding ab initio calculations
 allow in particular a quantitative description of the 
effect. Concerning this it could be demonstrated that an
 improved description of correlation effects on the basis
 of the LSDA+DMFT scheme leads to a substantially improved
 agreement of the theoretical and experimental spectra. This implies
 that achievements made for other systems by use of the DMFT can be
 monitored by our new combined approach in a most stringent way by
 comparing calculated photoemission spectra, that include all matrix
 element effects, directly to corresponding experimental data.

This work was funded by the German 
BMBF (Bundesministerium f\"ur Bildung und Forschung)
under contract FKZ 05 KS1WMB/1.

%

%
%

\end{document}